\documentclass{aastex}
\begin{document}

\title{ROSAT and BeppoSAX evidence of soft X--ray excess emission 
in the Shapley supercluster:
A3571, A3558, A3560 and A3562}

\author{Massimiliano~Bonamente$\,^{1}$, Richard~Lieu$\,^{2}$, Jukka~Nevalainen$\,^{3}$
and Jelle~S.~Kaastra$\,^{4}$}

\affil{\(^{\scriptstyle 1} \){Osservatorio Astrofisico di Catania, Via S. Sofia
78, I-95125
Catania, Italy}\\
\(^{\scriptstyle 3} \){Department of Physics, University of Alabama,
Huntsville, AL 35899, U.S.A.}\\
\(^{\scriptstyle 2} \){Astrophysics Division, ESA, ESTEC, Postbus 299, NL-2200 AG
Nordwijk, The Netherlands}\\
\(^{\scriptstyle 4} \){SRON Laboratory for Space Research, Sorbonnelaan 2, 3584 CA Utrecht, 
The Netherlands}\\
}

\begin{abstract}
Excess soft X--ray emission in clusters of galaxies has so far been detected 
for sources that lie along lines--of--sight to 
very low Galactic HI column density (such as Coma,
A1795, A2199 and Virgo, $N_H \sim 0.9-2.0 \times 10^{20}$ cm$^{-2}$). 
We show that the cluster {\it soft excess} emission 
can be investigated even at higher $N_H$,
which provides an opportunity for investigating soft X--ray
emission characteristics among a large number of clusters.
 
The ROSAT PSPC analysis of some members of the Shapley concentration
(A3571, A3558, A3560 and A3562, at $N_H \sim 4-4.5 \times 10^{20}$ cm$^{-2}$) 
bears evidence for excess emission 
in the 1/4 keV band. We were able to confirm the finding
for the case of A3571 by a pointed SAX observation.
Within the current sample the soft X-ray flux is again found to
be consistently above the level expected from a hot virialized plasma.
The data quality is however insufficient to enable a discrimination between
alternative models of the excess low energy flux.

\end{abstract}

\section{Introduction}
Clusters of galaxies are strong emitters of soft X--rays, often in excess
of the contribution from the hot intra--cluster medium (ICM; Lieu at al. 1196a;
Lieu et al. 1996b; Mittaz, Lieu and Lockman 1998; Lieu, Bonamente 
and Mittaz 1999;
Bonamente, Lieu and Mittaz 2001). The PSPC detector onboard
ROSAT is well suited for the observation of diffuse sources in the
1/4 keV band: along with a large effective area (of about 200 cm$^2$ at 0.28 keV),
the large field-of-view of 1 degree radius allowed the contemporaneous acquisition
of source signals (usually near boresight) and local background (at large detector radii)
with equal exposures. The Low--Energy Concentrator Spectrometer (LECS) aboard BeppoSAX 
also has appreciable
effective areas (of $\sim 10$ cm$^{2}$ at 0.28 keV)
and can be used for similar investigations.

Soft X--ray fluxes are significantly absorbed by intervening Galactic material 
(mainly HI and HeI) along the
line of sight.
At the Galactic N$_H$ values of the Shapley supercluster 
($\sim 4-4.5 \times 10^{20}$ cm$^{-2}$),
0.2 keV photons encounter 
a total optical depth of $\sim 4$,
which accordingly reduces
their flux by about 98 \%.
The HeI absorption cross--section was recently revised
(Yan et al. 1998 and
references therein), leading to a reinstatement of the original
absorption code of Morrison and McCammon (1983) as the most
reliable.  This code is therefore employed
in the present work.  The crucial task, however, is to measure $N_H$ with a
spatial resolution $\leq$ the cluster size.  
For details on how this was fulfilled, see section 2.
In this paper we present results
on a sample of nearby clusters (A3571, A3558, A3560 and A3562, at
redshift z$\sim$ 0.04-0.05)
 in the Shapley concentration. The analysis of four  PSPC observations 
revealed statistically
significant 1/4 keV band ($\sim$ 0.2--0.4 keV) 
excess emission for {\it all} of the clusters, 
although the intrinsic low surface brightness
of A3560 and A3562 implied a limited S/N.   Data from two pointed
SAX/LECS observations were available for two
of the four clusters (A3571 and A3562);
clear evidence of soft excess emission is present in the A3571 data, while
results for A3562 were marginal, due to the faintness of the cluster.
Thus
the soft excess syndrome may {\it be common to a large fraction of clusters}, as
suggested also by our earlier analyses of a nearby sample of clusters with
{\it EUVE}, contrary to the findings of Bowyer et al. (1999,2001)
which claim no general importance of the effect (see also
 Lieu et al. 1999 and Bonamente, Lieu 
and Mittaz 2001 for reinforcement of the original conclusions through confirmatory
re-observations conducted in such modes as to address all the objections of our
critics.).
This {\it paper} presents the data analysis method and
results for the current sample, with emphasis on the PSPC analysis; 
future works will be devoted to larger scale surveys, 
with both PSPC and LECS instruments, 
in order to assess the cosmological impact of the phenomenon. 

\section{Infrared, 21--cm and X--ray observations of A3571, A3558, A3562 and A3560}

The Shapley Supercluster is a large region of galaxy overdensity around $\alpha=13^h 25^m$,
$\delta=-30^o$ and redshift $\sim$ 0.045 (Shapley 1930; Ettori, Fabian and White 1997 and
references therein) comprising more than 20 galaxy clusters.
We  focused our attention on four Supercluster members: A3571 and A3558 are rich and hot clusters
(Markevitch et al. 1998;  Markevitch and Vikhlinin 1997), 
A3560 (Ebeling et al. 1996)
 and A3562 (White 2000; see also Ettori et al. 2000) 
are within $\sim$ 2 degrees 
from the former and show lower X--ray
luminosity (Fig. 1). All clusters were targets of pointed PSPC observations 
(archival identification code RP800287 for A3571, exp. time 5 ks; RP800076 for A3558, 28 ks; 
RP800381 for A3560, 5 ks;
RP800237 for A3562, 18 ks), and two of them of pointed SAX LECS/MECS observation 
(observation number
60843002 for A3571, 29 ks of LECS exposure, and 60638001 for A3562, 19 ks) .

Dedicated narrow--beam 21--cm observations of galaxy  clusters are 
available from the NRAO 100--meter GBT telescope
\footnote{The National Radio Observatory is operated by Associate Universities,
Inc.,
under cooperative agreement with the National Science Foundation.} with
a resolution of 10 arcmin
(Murphy, Sembach and Lockman 2001; see Table 1).  In addition
we also consulted the Dickey and Lockman (1990) survey data at lower resolution and
IRAS 100 $\mu$m maps with $\sim$ 3 arcmin resolution (Wheelock et al. 1994).
In particular, the IRAS data has a two--fold relevance to soft X--ray studies:
on one hand the positive N$_H$--IR correlation allows the use of 100 $\mu$m maps
as tracer of the total column density of Galactic material (Boulanger and 
Perault 1988);
on the other hand, the N$_H$--soft X--ray 
background (SXB) anticorrelation (Snowden et al. 1998)
 requires precise knowledge of the HI 
column density at the position where the local
background was determined.  In the case of PSPC this is
typically a few $\times$ 10 arcmin away from the cluster center.

\section{PSPC analysis}

The analysis of PSPC data was performed following 
the prescriptions of Snowden et al. (1994); in particular, 
detector gain corrections were applied using the dedicated FTOOL PCPICOR, and 
detector vignetting was accounted for in the background subtraction procedure. The spectral analysis
involved only PI channels 20--201 ($\sim$ 0.2--2.0 keV by photon energy), in order
to avoid any calibration uncertainties of the lower energy 1--19 PI channels.
Following the
nomenclature reported in Snowden et al. (1994), PI channels 20-41 will be referred to as the
R2 band (also 1/4 keV or C band), and channels 42--69 as R34 band. 

First, we compared the IR emission of the cluster regions
with the emission in the corresponding  background regions (Fig. 1);
deviations were always within at most 10 \%, and this implies that
the SXB of the background regions is indeed relevant to the cluster regions without
further scaling (see Snowden et al. 1998).
Secondly, a comparison of the IR emission between the cluster regions covered by the
GBT beam ($\sim$ 10 arcmin angular diameter) 
and the outer cluster regions showed that
the N$_H$ is smoothly distributed on the cluster's scale. In accordance to
the N$_H$--IR correlation (with a scaling factor of about 1.2 $\times 10^{20}$
cm$^{-2}$ (MJy sr$^{-1}$)$^{-1}$; Boulanger and Perault 1988) 
the 100 $\mu$m maps
implied only marginal adjustments from the central N$_H$ measurements; as a consequence,
we decided to apply Galactic column densities as in Table 1 for all subsequent analyses.

The limited passband of the PSPC instrument can only be used to measure
approximate values of the cluster hot gas temperatures.
Thus we modelled the A3571 and A3558 spectra
for annular regions with the best--fit temperatures of ASCA (Markevitch et 
al. 1998; Markevitch and Vikhlinin 1997; in agreement with SAX/MECS 
results for A3571, see 
Nevalainen et al. 2001), with the exception of the central regions where
temperature gradients are too steep for the point spread 
function (PSF) of ASCA: here
$T$ was fitted to the data. Temperatures of A3560 and A3562 were
determined by spectral fits to the PSPC data, and results 
agreed with previously published measurements 
(White 2000 and Ebeling et al. 1996). 
We modelled the entire PSPC X--ray spectrum (0.2--2.0 keV) 
with a one--temperature photoabsorbed thin--plasma
emission code (WABS and MEKAL codes in XSPEC).

Background spectra for subtraction purposes were extracted from same pointed
observation as the cluster's, and from annular regions 
immediately outside the detected
cluster emission (see Fig. 1); 
statistical errors in the background were properly 
propagated. This provides a spatially and temporally contiguous
(or {\it in situ}) background.
Best parameter values for the hot gas are shown in Table 1.
As the $\chi^2$ values indicate, some of the fits are poor,
due to the excess emission above the 1-T model at low energies. In more detail,
we compared
the best--fit model with the measured emission in the R2 and R34 bands
(Fig. 2 and 3). The 1/4 keV band results show consistently  {\it positive} 
residuals for all of the 4 clusters; the R34 band bears, on the other hand,
evidence of depleted fluxes in some regions.
For A3571 and A3558, there is evidence of an increasing relative importance
of the excess component with cluster radius, an effect already found
in EUVE data of other galaxy clusters (Virgo, A1795 and A2199; see, e.g.,
Bonamente, Lieu and Mittaz 2001 and Lieu, Bonamente and Mittaz 1999).
    
\section{SAX LECS analysis}

The SAX mission lends an additional view of the phenomenon,
with an instrument, LECS, 
that employs different and complementary
characteristics. With a field of view of radius $\sim 16$ arcmin, 
local background can be obtained only for compact clusters
(such as A3562); a long-term campaign of accumulation of high Galactic latitude
`blank' fields, however, enables one to establish a
template detector background  applicable to clusters which
fill the entire FOV (such as A3571; see Parmar et al. 1999). 
The energy dependent Point Spread Function (PSF) of the LECS instruments
complicates the analysis of extended objects such as clusters of galaxies, as
its 90\% encircled energy radius increases from 6\arcmin\ to
9\arcmin\ between 8~keV and 0.2~keV. In this work, we accounted for the PSF by
generating appropriate instrument response files to be used when fitting the
accumulated spectra. For the response generation we used the ray-tracing code
LEMAT (Lammers 1997), taking into account the extraction region size and
source position on the detector, mirror vignetting and obscuration by the
detector support structure (strongback).~\footnote{
 For any given spatial region and energy range, LEMAT estimates the
contribution of photons from regions external to the extraction region, and
the number of photons originating from within the extraction region that are
detected externally (see Kaastra et al. (1999); Nevalainen et al. (2001) for
the application of LEMAT in the case of the clusters A2199 and A3571).
Further applications of LEMAT to spectra of extended sources is in progress
(Oosterbroek 2001); preliminary results indicate that using 
the LEMAT--produced responses in spectral fits of spectra obtained within a 
radius of 15' removes the significant instrumental distortions (except for the 
whole band normalization which can vary by $\pm$ 30\%).}

Modeling
of the low-energy emission from A3571 was performed through a joint fit
of MECS/LECS spectra to a photoabsorbed MEKAL code 
(as for PSPC), which returned the
best-fit parameters of the hot ICM consistent with those of Table 2 (see
Nevalainen et al. 2001 for the details on the LECS/MECS analysis
of this data set):
excess emission in the 0.15-0.3 keV region\footnote{At 0.28 keV, the LECS energy
resolution
is $\sim$ 32\% (Parmar et al. 1997), about 3 times better than PSPC's. This
renders it difficult to translate PSPC pulse-invariant (PI) channels 
into corresponding LECS band for
the purpose of comparison. Our present choice of using 0.15-0.3 keV band 
of LECS as {\it equivalent} to PSPC PI channels 20-41 is based on the 10\% peak
response studies of Snowden et al. (1994). Similarly, PSPC R34 band has 10 \% of peak
response 0.2--1.01 keV;
moreover LECS data of A3571 also bear evidence of depleted fluxes around 
0.5 keV (Nevalainen et al. 2001), though no attempt is made here to extract
and compare such fluxes with those of the PSPC.}
is again detected with high statistical
significance (Fig. 4).
The faint emission from the A3562 cluster was 
likewise assessed by modelling LECS spectra
with the customary  photoabsorbed MEKAL code, where the hot ICM parameters were
fixed at the values of Table 2; results are marginalized by
the lower S/N (see Fig. 4).
It is worthwhile to point out that the {\it in situ} background, available
for this cluster from an outer annulus
where the source emission ceased, is fully consistent with the template
'blank' field background.  This is a crucial test which verifies the
reliability of the latter for those cases where the size of the cluster
renders the former unavailable.

\section{Interpretation of R2 and R34 band fluxes}

The results shown in Fig. 2 (and Fig. 4) indicate that the 1--T model employed to describe the ICM
falls short of the detected soft emission, revealing the {\it soft excess} emission.  
The data quality for A3571 and A3558 warrants the employment of a more
complex model to try to interpret the detected spectral features. 

The A3558 soft excess is virtually unaccompanied by a R34 trough, 
and therefore a second optically--thin
plasma component at lower temperature can model the excess. 
We shall present the 3--6 and 6--9 arcmin 
annuli as representative regions, just
outside the central cooling flow, and where signals are of sufficient statistical
quality to apply detailed modelling, see Table 2. 
We likewise tested
the possiblity of the excess originating as an inverse Compton (IC) effect, 
with the result that the data
equally well accomodates the non--thermal interpretation 
(see Table 2 for details). We also 
calculated the energetic requirements of the additional model, with the results that 
the luminosity of either thermal/non thermal 
component accounts for only a fraction of the
hot ICM luminosity in the 0.1--0.4 keV band.   

Moving to A3571, we again describe  a
representative region, the 4--8 arcmin annulus.
Here the $\sim 3 \sigma$ R2 band excess
pairs with a depleted emission in the neighboring R34 band (Fig. 2 and 3.), i.e. the employement of
a second thin plasma model at lower T brings only a modest improvement to the fit.
We know, however, that warm photoionized gas (T $\sim 10^6$ K) may present substantial
opacity to soft X--ray photons (e.g., Krolik and Kallman 1984), 
and the thin--plasma approximation may not necessarily apply.
A phenomenological way of parametrizing such behaviour is the addition of a second thin--plasma 
model modified by an absorption edge, which in fact
proves a successful fit to the data (see best--fit parameters in Table 2).
Other regions of Table 2 show a qualitatively similar behavior, see Fig. 2 and 3;
statistical quality of the data does not however warrant a more detailed modelling for them.
 
We also entertained the notion of a soft X--ray excess 
as peculiar variations of the Galactic N$_H$ (Arabadjis and
Bregman 1999), by applying
the same 1--T thermal models above leaving the N$_H$ to be fitted by the data. 
The A3571 and A3558 PSPC spectra would in such case require a Galactic column density
at odds with the measured values, and some 1/4 keV band excess would nonetheless
remain unexplained (see Table 1 and Fig. 5).    

\section{Discussion and conclusions}

The analysis we present in this paper 
shows that, at soft X--ray energies, PSPC spectra  of the clusters in
our sample (A3571, A2558, A3560 and A3562) are {\it not} well described 
by a simple one--temperature thin plasma model. 
Similar conclusions hold for the LECS data
of A3571 and A3562.
The case of A3558, the cluster with the higest S/N data of the sample,
 indicates that the soft excess can be interpretated as having a
thermal origin (from a second phase 
of the ICM at lower temperature) as well as the non--thermal.  
Earlier {\it EUVE} observations led to the discovery of soft excess emission
for the brightest clusters with low Galactic HI columns (such as Coma 
and Virgo, see references in section 1), but they lack the
necessary sensitivity to enable a reliable assesment of fainter sources, 
such as those reported in this work.
The PSPC and LECS data presented here show that, given a sufficient
sensitivity, low energy spectra of galaxy clusters {\it consistently}
 show excess
emission above the hot ICM contribution, in sharp contrast with the
conclusions of Bowyer et al. (2001). 
Settlement of the issue regarding the
origin of the emission has to
await forthcoming observations by detectors of higher
 resolution.

\begin{acknowledgements}
The BeppoSAX satellite is a joint Italian-Dutch programme.
We thank the staff of the BeppoSAX Science Data and
Operations Control Centers for help with these observations.
MB and RL acknowledge the support of NASA/ADP.
JN acknowledges an ESA Research Fellowship.
\end{acknowledgements}

\newpage 
\section*{References}

\noindent
Arabadjis,J.S. and Bregman, J.N. 1999, {\it ApJ}, {\bf 514}, 607. \\
\noindent 
Bonamente, M., Lieu, R. and Mittaz, J.P.D. 2001, {\it ApJ} {\bf 547}, L7. \\
\noindent
Bowyer, S., Bergh\"{o}fer, T.W. and Korpela, E.J. 1999, {\it ApJ}, {\bf 526}, 592.\\
\noindent
Bowyer, S., Korpela, E.J. and Bergh\"{o}fer 2001, {\it ApJ}, {\bf 548} L135. \\
\noindent 
Boulanger, F. and Perault, M. 1988, {\it ApJ}, {\bf 330}, 964. \\
\noindent
Dickey., J.M. and Lockman, F.J. 1990, {\it Ann. R. Astron. Astrop.}, 
{\bf 28}, 215. \\
\noindent
Ebeling, H., Voges, W., B\"{o}ringher, H., Edge, A.C., Huchra, J.P. 
and Briel, U.G. 1996, \\
\indent  {\it MNRAS}, {\bf 281}, 799. \\
\noindent
Ettori, S., Fabian, A.C. and White, D.A. 1997, {\it MNRAS}, {\bf 289}, 787. \\
\noindent
Ettori, S., Bardelli, S., De Grandi, S., Molendi, S., Zamorani, G.
 and Zucca, E. 2000, \\
\indent {\it MNRAS}, {\bf 318}, 239. \\
\noindent
Kaastra, J.S., Lieu, R., Mittaz J.P.D. et al., 1999, {\it ApJ}, {\bf 519},
119.\\
\noindent
Lammers, U. 1997, {\it The SAX LECS Data Analysis System - Software User \\
\indent Manual} , ESA/SSD, SAX/LEDA/0010. \\
\noindent
Lieu, R., Mittaz, J.P.D., Bowyer, S., Lockman, F.J., Hwang, C.-Y. and \\
\indent Schmitt, J.H.H.M. 1996a, ApJ, {\bf 458}, L5.\\
\noindent
 Lieu, R., Mittaz, J.P.D., Bowyer, S., Breen, J.O.,
Lockman, F.J., \\
\indent Murphy, E.M. \& Hwang, C. -Y. 1996b, {\it Science}, {\bf 274},1335--1338
. \\
\noindent Lieu, R., Bonamente, M. and Mittaz, J.P.D. 1999,
{\it ApJ}, {\bf 517}, L91.\\
\noindent
Lieu, R., Ip, W.-I., Axford, W.I. and Bonamente, M. 1999, {\it ApJ}, {\bf 510}, L25.\\
\noindent
Markevitch, M. et al., 1998, {\it ApJ}, {\bf 503}, 77. \\
\noindent
Markevitch, M. and Vikhlinin, A., 1997, {\it ApJ}, {\bf 474}, 84. \\
\noindent
~Mittaz, J.P.D., Lieu, R., Lockman, F.J. 1998, {\it ApJ}, {\bf
498},
L17. \\
\noindent
 Morrison, R. and McCammon, D. 1983, {\it ApJ}, {\bf 270}, 119.\\
\noindent
Murphy, E.M., Sembach, K.R. and Lockman, F.J. 2001, {\it in preparation}. \\
 \noindent Yan, M., Sadeghpour, H.R. and Dalgarno, A. 1998, {\it ApJ}, {\bf 496},
 1044.\\
\noindent
Nevalainen, J., Kaastra, J.S. and Parmar, A.N. 2001, {\it A \& A in press}. \\
\noindent
Oosterbroek, T. et al. 2001, {\it in preparation}.\\
\noindent
Parmar, A.N. et al. 1997, {\it A \& AS}, {\bf 122}, 309.\\
\noindent
Parmar, A.N et al. 1999, {\it A \& AS}, {\bf 136}, 407.\\
\noindent
Shapley, H. 1930, {\it Harvard Obs. Bullettin}, {\bf 874}, 9. \\
\noindent Snowden, S.L., McCammon, D., Burrows, D.N. and Mendenhall, J.A. 1994,
 {\it ApJ}, {\bf 424}, 714. \\
\noindent Snowden, S.L., Egger, R., Finkbeiner D.P., Freyberg, M.J. and  \\
\indent Plucinsky, P.P. 1998, {\it ApJ}, {\bf 493}, 715.\\
\noindent
Wheelock, S. et al., 1994, {\it IRAS Sky Survey Expl. Supp.}, (IPL Publ. 94-11),
Pasadena, CA. \\
\noindent
White, D.A. 2000, {\it MNRAS}, {\bf 312}, 663. \\
\noindent Yan, M., Sadeghpour, H.R. and Dalgarno, A. 1998, {\it ApJ}, {\bf 496},
 1044.\\

\newpage
\section*{Figure captions}
Figure 1: Schematic
of the field of the four clusters (A3571, A3558, A3560 and A3562).
Large solid circles represent the approximate cluster positions,
the two small circles between A3562 and A3558 
are the groups of galaxies SC1329 and A1327
(respectively 
from left to right; see Ettori et al. 1997 for more detailed maps).
Each dotted annulus (or sector thereof) represents the approximate
 region from which the background 
for the corresponding cluster was extracted; care was taken in avoiding regions
occupied by SC1329 and SC1327 and obvious point sources. 

Figure 2: Radial distribution (in arcmin, x-axis) of the
fractional R2 band excess, $\eta$, of the four clusters,
defined as $\eta = (p_2-q_2) /q_2$, where $p_2$ is observed R2 band flux 
and $q_2$ the expected R2
band flux from the hot ICM.  Negative values of $\zeta$ indicate
that detected emission is below the expected value.

Figure 3:  As in Figure 2, except now for the
fractional R34 band excess
$\zeta = (p_{34}-q_{34})/q_{34}$ where $p_{34}$ is observed R34 band flux and
$q_{34}$ is expected R34 band flux from hot ICM.

Figure 4: Radial distribution of
the SAX/LECS fractional excess in the 0.15-0.3 keV band for
A3571 and A3562, plotted in the same manner as Figure 2.


Figure 5: As in Figure 2, except here the expected
ICM fluxes are derived by fitting $N_H$. Soft excess fluxes in some regions of A3571
remain nonetheless unexplained. In order to eliminate such excesses,
we estimate that a $N_H$ of 4.05 $\pm^{0.15}_{0.2} \times 10^{20}$ (4-8 arcmin region) and 
3 $\pm^{0.5}_{0.4} \times 10^{20}$ (12-16 arcmin region) would be required, 
significantly {\it lower} than the measured values (see Table 1).

\newpage
 
\section*{Tables}
\begin{table}[h!]
\begin{center}
\tighten
\caption{Best parameters of the single temperature
hot gas for the clusters in the sample.
Errors are 90 \% confidence level ($\chi^2$ + 2.7 criterion); where no
errors are reported, the parameters were not fitted to the data (see text). Fitted T values
are obtained keeping NH fixed at the Murphy, Sembach and Lockman (2001) values.
In bold-face are the 
$N_H$ values from pointed observations of Murphy, Sembach and Lockman (2001);
the survey values are from Dickey and Lockman (1990), 
with average over a 3$\times$3 square
degrees area.
\label{tab:tab2}}
\vspace{22pt}
\hspace{-2cm}
\small
\begin{tabular}{cccccccc}
\hline
 Cluster & region (arcmin) &N$_H$ & Survey N$_H$& best-fit N$_H$& T (keV) &
  A & $\chi^2$ (d.o.f) \\
\hline
A3571 & 0--2~\tablenotemark{a} & {\bf 4.4} & 3.9 & 3.77 $\pm^{0.47}_{0.37}$ & 4.5 $\pm^{2.0}_{1.1}$ & 0.6$\pm^{0.61}_{0.38}$ &113(139)\\
      & 2--4 & " & " & 4.54 $\pm^{0.33}_{2.9}$ & 7   & 0.3 & 159(139)\\
      & 4--8 & " & " & 4.36 $\pm^{0.36}_{0.3}$ & 7   & 0.3 & 233(162) \\
      & 8--12& " & " & 4.12 $\pm^{0.78}_{0.74}$ & 7   & 0.3 & 98(116)\\
      & 12--16 & "  & " & 4.25 $\pm^{3}_{1.25}$ & 5 & 0.3 & 110(92)\\
\hline
A3558 & 0--3  & {\bf 4.0} & 3.9 & 3.75 $\pm^{0.21}_{0.2}$ & 3.2 $\pm^{0.4}_{0.35}$ & 0.33 $\pm^{0.1}_{0.1}$ & 209.6(179)\\
      & 3--6  & {\bf 4.0} & " & 3.65 $\pm^{0.15}_{0.15}$ & 5     & 0.3 & 211(180)\\
      & 6--9  & {\bf 4.0} & " & 3.4 $\pm^{0.17}_{0.2}$ & 5     & 0.3 & 237.7(176)\\
      & 9--12 & 4.1 & " & 3.5 $\pm^{0.37}_{0.38}$ & 5     & 0.3 & 158(169)\\
      & 12--15& 4.2 & " & 3.1 $\pm^{0.57}_{0.55}$ & 5     & 0.3 & 161.6(158)\\
      & 15--18& 4.2 & " & 3 $\pm^{1}_{0.75}$  &3.5   & 0.3 & 27(19)\\
\hline
A3562 & 0--3 & {\bf 4.0} & 3.85 &4 $\pm^{0.26}_{0.33}$ & 2.7 $\pm^{0.5}_{0.3}$ & 0.6&203.6(165) \\
      & 3--6  & " & " & 3.56 $\pm^{0.5}_{0.35}$ & 4.5 $\pm^{1.9}_{1.1}$ & 0.3& 134(161) \\
      & 6--12 & " & " & 3.65 $\pm^{0.65}_{0.5}$ & 3.6 $\pm^{1.2}_{0.8}$ & 0.3&205.4(163) \\
\hline
A3560 & 0--3 & {\bf 4.7} & 4.6 & 5.3 $\pm^{2.1}_{1.3}$ & 2.2 $\pm^{1.5}_{0.6}$ 
                   & 0.38$\pm^{0.62}_{0.25}$ & 52.7(47) \\
      & 3--6 & " & " & 4.2 $\pm^{2.5}_{1.6}$ & 4.2 $\pm^{10}_{2.15}$ & 0.3 & 85(58) \\
      & 6--12& " & " &  4.1 $\pm^{2.9}_{1.8}$ & 1.9$\pm^{1.0}_{0.5}$  & 0.25$\pm^{0.65}_{0.24}$ & 104.4(78) \\
\hline
\end{tabular}
\begin{flushleft}
\tighten
{\small a -- Use of best--fit values kT=7.5, A=0.5 (Markevitch et al. 1998)
will result in no changes to Fig. 2 and Fig. 3.}
\end{flushleft}
\end{center}
\end{table}

\begin{table}[h!]
\begin{center}
\tighten
\caption{Modelling the PSPC spectra of some regions of A3571 and A3558 with 
an additional thermal
or non--thermal component; parameters of the hot ICM were fixed 
at the best values as given in 
Table 2 except  the normalization constant (emission measure)
which was fitted to the data. For A3558, a second MEKAL
component is added to the model, with the same elemental abundances as those of the hot phase;
alternatively, a power--law model is added with its 
differential photon number index fixed at the value of $\alpha$=1.75,
corresponding to emitting electrons having the
Galactic index of cosmic rays (see, e.g., Lieu et al. 1999).  
For A3571, the model employed was, in XSPEC language, 
a WABS*EDGE*(MEKAL+MEKAL).  This
model is only intended as a preliminary means of accounting for
the observed spectral features,
as a model that includes details of radiative transport and opacity 
is not currently available. These spectral models were also satisfactorily applied
to the LECS data, showing that PSPC and LECS give consistent results.
\label{tab:tab3}}
\vspace{22pt}
\hspace{-2cm}
\small
\begin{tabular}{cc|cccc|c}
\hline
Cluster & region & \multicolumn{4}{c}{2$^{nd}$ thermal comp.} & 
2$^{nd}$ non--thermal comp. \\ 
        & (arcmin) & \multicolumn{4}{c}{ } &  \\
\hline
        &    & T (keV) & & & $\Delta \chi^2$ ($\Delta$ d.o.f.)&$\Delta \chi^2$($\Delta$ d.o.f.) \\ 
\hline
A3558   & 3--6 & 0.04 $\pm^{0.06}_{0.02}$ & & & 15(2) & 13(1) \\
        & 6--9 & 0.05 $\pm^{0.1}_{0.03}$ & & & 24(2) & 35(1) \\
\hline
        &    & T (keV) & edge E (keV) & edge $\tau$ & $\Delta \chi^2$ ($\Delta$ d.o.f.) &  \\ 
\hline
A3571   & 4--8 & 0.85 $\pm^{0.6}_{0.2}$ & 0.52$\pm^{0.04}_{0.09}$ & 0.98$\pm^{0.65}_{0.05}$& 51(4) & \\
\hline
\end{tabular}
\end{center}
\end{table}

\end{document}